\documentclass[twocolumn,amssymb,prd]{revtex4}

\usepackage{graphicx}
\usepackage{dcolumn}
\usepackage{bm}
\usepackage{epsfig}

\begin{document}
\newcommand{\gsim}{\hbox{\rlap{$^>$}$_\sim$}}
\newcommand{\lsim}{\hbox{\rlap{$^<$}$_\sim$}}

\title{Comment on AMS02 results support the secondary origin of cosmic 
ray positrons}

\author{Shlomo Dado} \affiliation{Department of Physics, Technion,
Haifa 32000, Israel}

\author{Arnon Dar} \affiliation{Department of Physics,
Technion, Haifa 32000, Israel}

\maketitle

Recently Blum, Katz and Waxman [1] have claimed that the flux of high 
energy cosmic ray (CR) positrons near Earth that has been measured with 
the Alpha 
Magnetic Spectrometer (AMS) [2] aboard the International Space Station 
can be produced in the collisions of Galactic CR protons and nuclei with 
the ambient matter in the Galactic interstellar medium (ISM). Their 
claim was based on their  alleged "robust upper limit to the positron 
flux" produced in the Galactic ISM, which neglected the energy loss of 
${\rm e^+}$'s in the ISM 
and ignored the possibility that secondary production 
could take place mostly inside the cosmic ray sources [3].

In this comment, we present a simple calculation of the flux of secondary 
positrons produced in the ISM that is based only on priors. Our 
calculated ISM flux agrees very well with that calculated with the 
elaborate GALPROP code [4]. It confirms that secondary production of 
positrons in the ISM by the primary cosmic rays cannot explain the 
observed sub-TeV flux of CR positrons. Moreover, we show that once energy 
loss of positrons in source and in the ISM are included, secondary 
production inside the CR sources plus the ISM does explain the measured 
near-Earth flux of cosmic ray positrons.
  
The total flux of primary nucleons (n) in the energy range between 
several GeV and PeV 
is well described by  ${\rm \phi_n(E)\approx 1.8\, 
(E/GeV)^{-\beta}\,  GeV^{-1}\, cm^{-2}\, s^{-1}\, sr^{-1}}$
where ${\rm \beta =2.7}$. 
Nucleons bound in nuclei (A,Z) have a larger effective rigidity (R=AE/Z)
and a smaller effective cross section (${\rm 
\sim\sigma_p/A^{1/3}}$)  than free protons.
Hence, in a steady state situation, the secondary sub-TeV
${\rm e^+}$ flux from  
the decay of  ${\rm \pi^+}$'s and ${\rm K^+}$'s produced in hadronic 
collisions between the primary cosmic ray nucleons 
and ambient ISM matter is bounded by
\begin{equation}
{\rm \phi_{e^+}(E)\leq f_{e^+}\,\sigma_{in}(pp)\,n_{ism}\,c\,
\tau\,\phi_n/(\beta-1)}\,
\end{equation}  
where Feynman  scaling for CR nucleons with a spectral index 
${\rm\beta =2.7}$ yields ${f_{e^+}\approx 7\times 10^{-3}}$.
Secondary ${\rm e^+}$'s with energy in the range
10-100 Gev are produced mainly by
CR nucleons with energy in the range 100-1000 GeV, where
${\rm \sigma_{in}(pp)\approx 35}$ mb. The  mean 
baryon density ${\rm n_{ism}}$ in the Galactic cosmic ray halo 
is  independent of the rigidity of cosmic rays
as long as their Larmor radius is much smaller than the 
scale height H
of the magnetic trapping region of cosmic rays  above the Galactic 
disk. For ${\rm H\approx 4 \, kpc}$ [5]
obtained from fitting cosmic ray data with the
best available numerical calculations of  Galactic propagation of cosmic 
rays [4], this is valid for ${\rm E_p< 10^9\, GeV}$. The surface  
density  ${\rm \Sigma\approx 10\, M_\odot/pc^2)}$ of the Milky Way disk
obtained [6] from  21 cm radio observations of HI plus synchrotron 
emission   
from HII yield
${\rm n_{ism} \approx  \Sigma/2\,H\, m_p\approx 0.05\, cm^{-3}}$.
The mean life time
${\rm \tau=E/(dE/dt)}$ of positron due to
escape (esc) of ${\rm e^+}$'s from the Galaxy by diffusion in its
turbulent magnetic fields and by radiative (rad) energy losses (due 
to inverse
Compton scattering of background photons and synchrotron radiation)
satisfies ${\rm \tau=\tau_{esc}\,\tau_{rad}/(\tau_{esc}+\tau_{rad})}$.
Synchrotron emission and inverse Compton scattering of background 
photons in the Thomson regime yield
\begin{equation}
{\rm \tau_{rad}\approx  3\,(m_e\,c^2)^2/
4\, \sigma_{_T}\, c\, U\, E \approx 10^{16}\, (E/GeV)^{-1}\, s}
\end{equation} 
where ${\rm \sigma_{_T}\approx}$0.66 barn 
and ${\rm U \approx 1\, eV}$ is  the  total energy density of the
magnetic field plus that of the 
diffuse radiation fields
(mainly star-light, far infrared radiation, and the
cosmic microwave  background).
The escape times of protons and ${\rm e^\pm}$'s 
by diffusion in the Galactic random magnetic fields 
with a Kolmogorov spectrum
are practically identical and satisfy ${\rm \tau_{esc} 
\approx  10^{15}\, (E/GeV)^{-1/3}} $ s.

Fig.1 compares the 
predicted flux of secondary positrons 
produced in the ISM as given by  Eqs. (1),(2) 
and their measured flux [2]. 
It  shows that secondary production in the  
ISM by the primary CRs  can explain neither the magnitude nor 
the spectrum  of the observed CR ${\rm e^+}$  flux above 10 GeV.
\begin{figure}[] 
\centering 
\epsfig{file=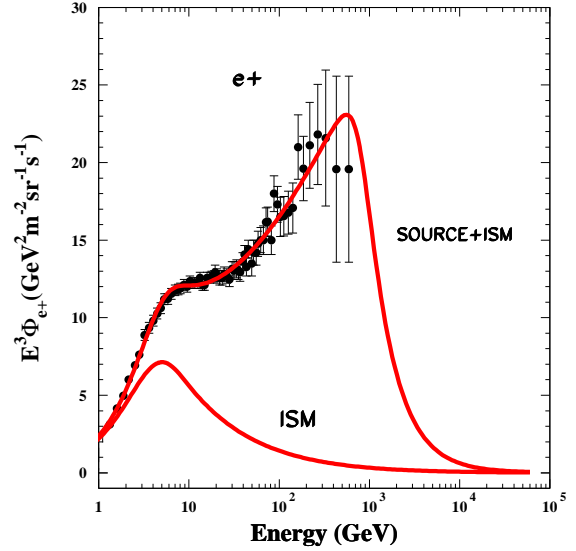,width=8.cm,height=8.cm} 
\caption{Comparison
between the CR positron flux measured with AMS [2] and
the flux expected from positron production in the ISM alone  and 
 in the ISM plus inside the Galactic CR sources. For more details see
arXiv:1505.04988.} 
\label{Fig1}
\end{figure}

The total flux of secondary ${\rm e^+}$'s produced 
inside source and in the ISM can also be obtained from Eq.~(1) 
by replacing ${\rm \,n_{ism}\,c\,\tau}$, 
the column density traversed by the cosmic ray nucleons in the ISM, 
with the total column density traversed  inside the CR 
sources and the ISM.
\begin{equation}  
{\rm \phi_{e^+}(E)\approx {f_{e^+}\,\sigma_{in}(pp)\,X\, \tau\, \phi_n
\over (\beta-1)\,m_p\, \tau_{esc}}}\,,
\end{equation}
where ${\rm X=8.7\, (E/10\, GeV)^{-0.4}\, g/cm^2}$  is the mean total 
grammage (column density times ${\rm m_p}$) traversed  by  high energy 
cosmic 
ray nucleons  with energy E per nucleon {\it inside the CR source and in 
the ISM}. In Fig. 1 we also compare the positron flux predicted by 
Eq.(3) for the total grammage 
extracted from the measured B/C ratio (bottom panel of Fig. 5 of Ref [1]).
The agreement is very good as can be seen from Fig.~1.

\end{document}